# On Automated, Continuous Chronotyping from a Daily Calendar using Machine Learning

Authors:

Pratiik Kaushik[1], Koorosh Askari[1], Saksham Gupta[2], Rahul Mohan[3], Kris Skrinak[4], Royan Kamyar, M.D.[2], Benjamin Smarr, Ph.D.[5]

1. Virginia Commonwealth University School of Medicine
2. Owaves, Inc.
3. Henry Samueli School of Engineering and Applied Science, University of California, Los Angeles
4. Amazon Web Services
5. Shu Chen - Gene Lay Department of Bioengineering and the Halicioglu Data Science Institute, University of California, San Diego

# Abstract

**Objectives:** Chronotypes - comparisons of individuals' circadian phase relative to others - can contextualize mental health risk assessments, as well as support the detection of social jet lag, which can hamper mental health and cognitive performance. Existing ways of determining chronotypes, such as Dim Light Melatonin Onset (DLMO) or the Morningness-Eveningness Questionnaire (MEQ), are limited by being discrete in time and time-intensive to update, meaning they rarely capture real-world variability across time. Chronotyping users based on a living schedule, such as a daily planner app, might augment existing methods to enable the assessment of chronotype and social jet lag continuously and at scale. Developing this functionality would require the use of a novel tool to translate between digital schedules and chronotypes. This paper reports the construction of a supervised binary classifier that attempts to demonstrate the feasibility of this approach.

**Methods:** In this study, 1,460 registered users from the Owaves app opted in by filling out the MEQ survey between July 14, 2022, and May 1, 2023. Of those, 142 met the eligibility criteria for data analysis. We used multimodal app data to assess the classification of individuals identified as morning and evening types from MEQ data, basing the classifier on app time series data. This includes daily timing for 8 main lifestyle activity categories (exercise, sleep, social interactions, meal times, relaxation, work, play, and miscellaneous) as defined in the app.

**Results:** The timing of activities showed substantial change across time, as well as heterogeneity of time by activity type. Our novel chronotyping classifier was able to predict the morningness and eveningness of its users with an ROC AUC of 0.70.

**Conclusion:** Our findings demonstrate the feasibility of chronotype classification from multimodal, real-world app data, while highlighting fundamental challenges to applying discrete and fixed labels to complex, dynamic, multimodal behaviors. Our findings suggest a potential for real-time monitoring of shifts in chronotype specific to different causes (i.e. types of activity), which could feasibly be used to support future, prospective mental health support research.

# Introduction

Circadian rhythms in humans manifest across a wide spectrum of biological and behavioral functions, encompassing hormone secretion, body temperature regulation, sleep-wake cycles, and social interactions. These intrinsic diurnal variations are governed by internal circadian clocks and modulated by external environmental cues, with light being a primary example. The expression of circadian rhythms varies significantly among individuals, influenced by genetic predispositions, environmental factors, age, and sex.[1–6]

Chronotype, a concept used to describe an individual's natural circadian pattern, is typically categorized into three types: morning, intermediate, and evening.[1,7–9] Morning-type individuals stereotypically exhibit a preference for earlier bedtimes and wake times, reaching peak cognitive and physical performance in the earlier in the day than other chronotypes. Conversely, evening-type individuals peak later in the day, favoring later sleep and wake schedules. Approximately 40% of adults are classified into one of the two extreme chronotypes, while the majority fall into the intermediate category (though this depends on the exact definition used in a particular study).[10]

One reason that chronotype is important is that it can make it easier to assess misalignment between an individual's usual daily rhythms and those of their environment. Increasing evidence suggests that circadian disruption may be a contributing cause of many chronic diseases, including but not limited to: obesity, type 2 diabetes, heart disease, depression, ADHD, and bipolar disorder.[8,11] The opportunity exists to improve or optimize circadian behaviors to achieve better outcomes in these conditions. To implement chronotherapy, for example, to advance or delay the patient's phase, a clinician, therapist, or coach may chronotype a patient to better understand their baseline behavior.[8,12] Circadian disruptions can be contextualized given the individual's baseline chronotype to provide successful therapeutic and coaching guidance.[12,13] Critically, an individual's chronotype may change over time, for example based on age, season, or location.[2,3] The goal of such therapies is therefore to modulate the patient's behavior longitudinally to support desired clinical outcomes.

Current methods for measuring chronotypes were not designed to support ongoing, frequent re-assessment. For example, dim-light melatonin onset (DLMO) requires repeat blood tests in a clinical setting.[14] While DLMO has long been a standard due to its superior stability in low-light conditions[15], that stability may make it less helpful for assessing the alignment across the many body-oscillators that compose an individual. The relative instability of other systems compared to DLMO supports the hypothesis that different body systems maintain different internal circadian phases, and respond to different environmental challenges at different rates. Additional markers could therefore augment DLMO if the goal is to have a physiological-network model across an individual's internal circadian rhythms. DLMO requires several hours in a clinical setting, and so the expense and burden make it usually not appropriate for longitudinal assessment of potential shifts in chronotype across time. Alternatives have been explored to predict DLMO from wearable sensors[16], and these hold promise, but there remain many contexts in which wearable sensors may not be available. Surveys have also emerged as a standard chronotype assessment tool, and these are cheap and easy enough to administer that they have very broad reach. Long-form surveys, such as the Morningness-Eveningness Questionnaire (MEQ) and Munich Chronotype Questionnaire (MCTQ), have the advantage of being taken from home without the need for additional equipment or cost. However, they are still limited due to the time and concentration required to fill out survey[17], as well as the discrete nature of those assessments - by design these surveys attempt to capture average circadian and sleep information, and so are not designed to capture change across days. Traditional survey instruments may therefore not be ideal tools during the type of chronotherapeutic intervention described above, as they do not provide continuous, ongoing insight into potential chronotype shifts of the patient.

An underexplored method for assessing chronotype is the utilization of an individual's digital real-world schedule, as obtained from a calendar or day planner software application. If an individual has sufficient freedom to reflect preferences into his/her/their schedule, it may be feasible to develop an algorithm that translates real-world calendaring data into an accurate chronotype prediction. This approach is supported by findings that online activity patterns can be used to extract chronotype information from individuals (e.g. [6,18,19]). This approach has the potential to capture more longitudinal data than traditional methods and allows for more passive participation if the participant was using the calendaring app for their own purposes already, as opposed to only for the purpose of supporting chronotype.

This study presents the development of a machine learning algorithm designed to test the feasibility of automatically, continuously, and accurately classifying an individual's chronotype utilizing data from the *Owaves: Body Clock Calendar* app.

# Methods

## Participants and Data Collection

Participants were drawn from a subset of Owaves app users who had previously agreed to share their data to support research during account registration. Registered app users were presented with the opportunity to complete the full Morningness-Eveningness Questionnaire (MEQ) via the following prompt:

*"Find out your Body Clock type! Your answers will help make our app smarter. If you have a registered Owaves account, we would like your permission to analyze your answers to this quiz. We would also like your permission to compare your quiz results with your existing Owaves account information. This will help us improve our ability to understand Body Clocks and provide intelligent recommendations and insights in the future. This is completely voluntary!"*

This was carried out internally at Owaves before the involvement of other researchers. For the purpose of this analysis, MEQ data were used if they were gathered between July 14, 2022, and May 11, 2023. The paired activity data were sourced retrospectively from the Owaves app, collected between December 1, 2017, and June 20, 2024. All data was aggregated and anonymized before analysis so that all research was carried out on pre-existing and de-identified data, and as such did not meet the criteria of research requiring an IRB. Pre-existing, de-identified data were provided by Owaves to its internal research team as an independent data set containing time stamps per activity category, as well as self-reported gender and age (to within two years to support anonymity). No location data aside from timezone sufficient to calculate local time, ethnicity data, IDs of any kind, or other potential sources of re-identification were included in the data set. All analyses were carried out appropriately to ensure privacy and confidentiality.

The inclusion criteria for participants were:
- Registered app user
- Age between 18 and 32 years
- Regular use of the Owaves app (≥14 days total; contiguity not required)
- Completion of the full MEQ survey

## Demographic Data

The demographic characteristics of the study participants are summarized in Table 1:

| Table 1: Demographic Data | | | | | | |
|---|---|---|---|---|---|---|
| Age Range | Count | Percentage | | Gender | Count | Percentage |
| 18-20 | 47 | 33.1% | | female | 105 | 73.9% |
| 21-23 | 33 | 23.2% | | non-binary | 19 | 13.4% |
| 24-26 | 27 | 19.0% | | male | 13 | 9.2% |
| 27-29 | 14 | 9.9% | | Prefer not to answer | 3 | 2.1% |
| 30-32 | 21 | 14.8% | | transgender | 2 | 1.4% |
| Total | 142 | | | Total | 142 | |

## Data Processing

The data from the Owaves app was filtered to remove duplicate events, and activities recorded before the year 2016. The MEQ scores were merged with the planned activities data and categorized as follows:

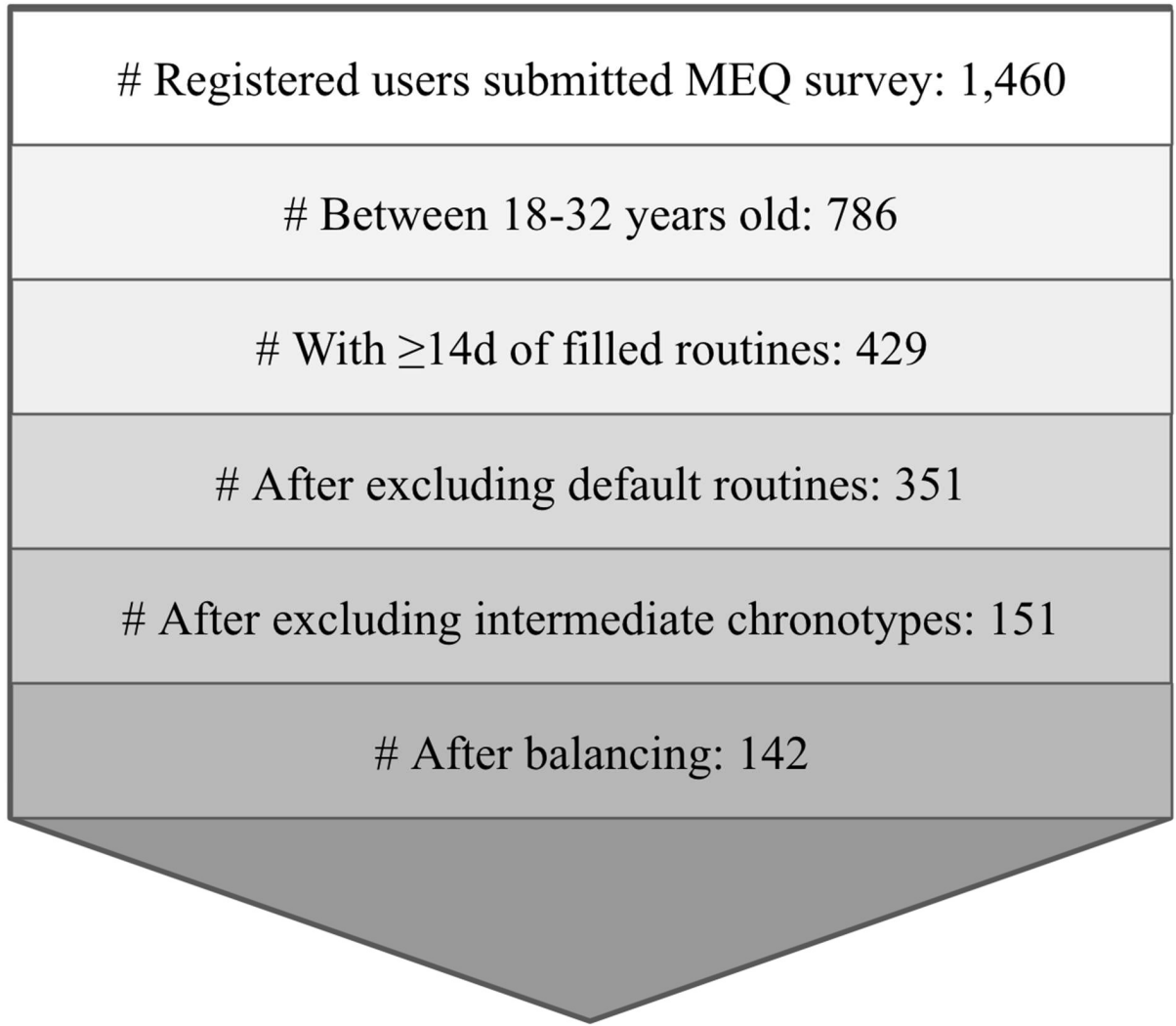

**Figure 1: Data attrition chart.** This shows the number of users processed within each cleaning step in the data pre-processing pipeline. Intermediate users are those with MEQ-derived "intermediate" chronotypes. See Data Processing for further details.

Further data transformation included removing rows with invalid start dates and converting the list of activities into a 96 column by number of rows matrix. Each column represented a 15-minute block of time, each row represented one user-day, and each cell contained the activity number for that specific time: -1: No activity, 0:Love/Relationships/Social, 1:Exercise/Physical Activity, 2:Work/Study/School, 3:Eat/Meals, 4:Sleep/Nap, 5:Relax/Mindfulness/Stress Management, 6:Play/Fun, 7:Flow/Miscellaneous/Errands. Formatting the data in this manner allowed for implicit encoding of time data - each column represented the same 15 minute block of the day for all users. It avoided two issues: the first of mixing data types in a column (by adding a row specifically for timestamps), and the second of having row length be a function of the number of activities a user planned in a day, an uncontrollable variable which resulted in a sparse input matrix unsuitable for machine learning.

The Owaves app provides two default routines ("Body Clock for Beginners" and "9 to 5 Alive") for new users with pre-specified activity types filling the 24h day. Existing users may instead customize their routines by filling their own activity information per each unit time. To avoid training on the default data which might appear identically across users of different actual chronotypes, days with over 85% identity to either of two default routines suggested by the app or more than 2.5 hours (10 15-minute blocks) without any activities were excluded to reduce the possible inclusion of default data instead of human-generated data.

For preprocessing, individuals with intermediate chronotypes (as determined by their MEQ) were removed from the analysis. Early and moderate morning types were merged into a single "morning" category, and early and moderate evening types were merged into a single "evening" category. User days were grouped into blocks of seven. The data was split into testing and training splits, ensuring that no user appeared in either of the splits. These two splits were balanced to ensure an equivalent number of user weeks from morning and evening types.

Principal Component Analysis (PCA) was performed on the midpoints between activity start and end times, with data points labeled according to the user's reported chronotype (Figure 2).

## Machine Learning Model

The classification algorithm was trained to differentiate between morning and evening chronotypes. The data was split into five folds, ensuring each user's data appeared in only one fold (all data per user were included per fold, to a minimum of 14 days). An XGBoost model was trained on these five-folds, with the F1 score (harmonic mean of precision and recall) used as the evaluation metric.

Hyperparameter optimization was conducted using the Syne-Tune HPO library. The search space for the hyperparameters was defined in Table 2:

| Table 2: Hyperparameter Optimization | | |
|---|---|---|
| **Hyperparameter** | **Range** | **Optimized Value** |
| Objective | - | binary |
| num_boosting_rounds | randint(64, 1024) | 98.00 |
| max_depth | randint(2, 8) | 6.00 |
| eta | loguniform(1e-3, 1) | 0.08 |
| gamma | loguniform(1e-6, 64) | 2.63 |
| min_child_weight | randint(0, 32) | 1.00 |
| subsample | uniform(0.5, 1.0) | 0.87 |
| colsample_bytree | uniform(0.3, 1.0) | 0.90 |
| reg_lambda | loguniform(1e-6, 2.0) | 0.06 |
| alpha | loguniform(1e-6, 2.0) | 0.002 |

The binary classifier was evaluated on 30 different subsamples of the data and aggregated, with the results displayed in a receiver operating characteristic (ROC) curve (Figure 4A) and a confusion matrix (Figure 4B).

# Results

## Demographics

Data preprocessing involved filtering out app-generated default routine data and transforming activity logs into a structured matrix format. This approach ensured that the classifier was trained on user-generated data. The exclusion of intermediate chronotypes and grouping of user days into weekly blocks refined the dataset for model training. Out of 1,460 users who completed the Morningness-Eveningness Questionnaire (MEQ) survey, 142 met the eligibility criteria. The majority were female (73.9%), with non-binary (13.4%) and male (9.2%) participants, primarily aged 18-20 years (33.1%).

## Data and PCA

We found that the timing of events is not stable across days. Activity types are also frequently not monomodal, but distributed in multiple non-contiguous events across the day. Figure 2 illustrates the daily activity profiles of one morning and one evening user over seven days, highlighting the variability and granularity in activity patterns.

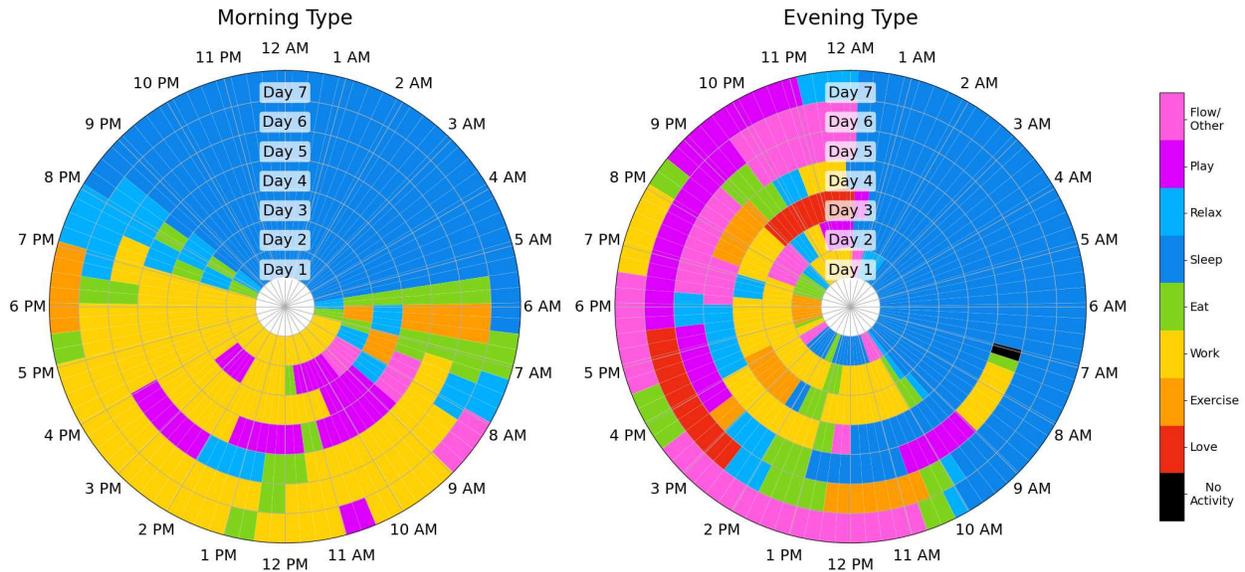

**Figure 2: Activity profiles change across days as well as across individuals.** One morning user's data (n=1; left) and one evening user's data (n=1; right), each over 7 consecutive days. No two days are exactly the same, reflecting the inherent variability in data used to assign chronotypes to people, and illustrating the granularity available through calendaring app data.

PCA conducted on the mean midpoints of each activity type per individual provides a means to visually assess the average separability of chronotypes by feature aggregate statistics – a rough approximation of the aggregation that must happen for the filling out of retrospective questionnaires like the MEQ. Figure 3 shows that while individual's midpoints projections showed some clustering by chronotypes, there was substantial overlap. There appeared to be some similarity between the daily activity patterns of people with different chronotypes, making discrete classification numerically difficult and inherently probabilistic. We calculated a Kullback-Leibler (KL) divergence score of 0.1669 in the PCA-transformed space. We then decided to test whether the use of individual data points, as opposed to aggregates, would improve the potential classification of chronotypes.

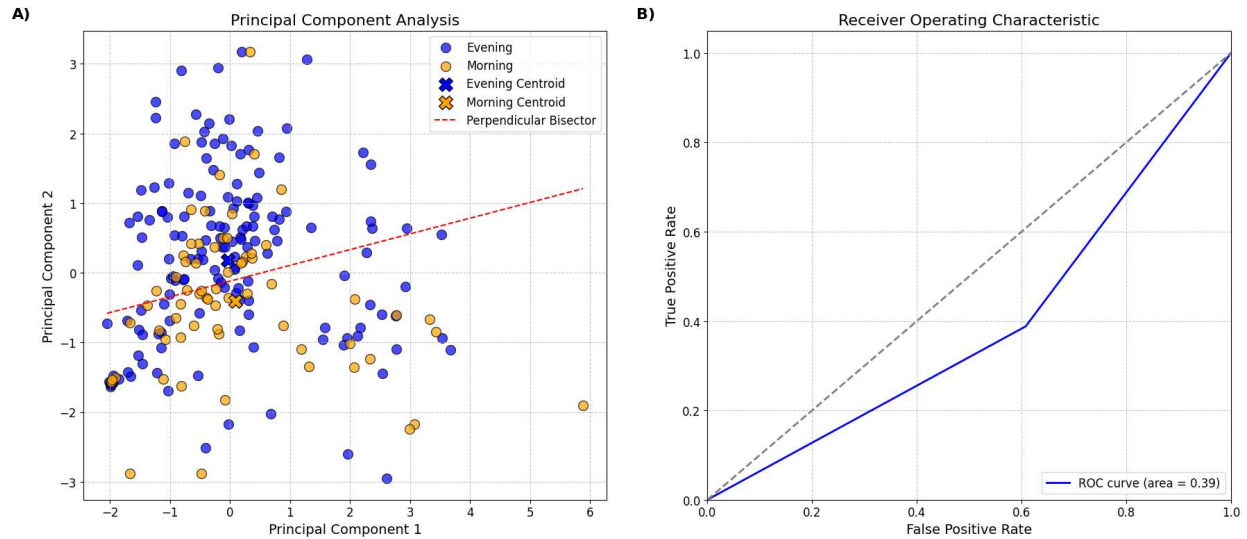

**Figure 3: PCA of activity data by chronotype reveals substantial overlap.** A) The scatter plot depicts the distribution of users' activity data (n=142) reduced first to mean daily midpoint per activity type, and then through PCA to two principal component dimensions (PCs) per individual. Morning chronotypes: orange; evening chronotypes: blue. An activity and time-naive classification approach involving this PCA and its performance is shown here. The centroid of the evening and morning populations were found, and the perpendicular bisector of the line connecting the centroids was used to classify each user. Users on the side closer to the morning centroid were classified as morning, and users on the side closer to the evening centroid were classified as evening. The area under the Receiver Operating Characteristic curve is shown in B), with an AUC of 0.39, slightly worse than random guessing.

## Machine learning

We trained an Extreme Gradient Boosting (XGBoost) binary classifier using five-fold cross-validation, achieving a Receiver Operating Characteristic (ROC) Area Under the Curve (AUC) of 0.70 with a standard deviation of ±0.058 (Figure 4A). The confusion matrix (Figure 4B) reveals that the classifier can successfully find true positives most of the time. The model is more accurate when classifying Evening types (lower % false positive) than when classifying Morning types, but in both cases true positive is the most likely outcome. When the column order is randomized within person before training, then the timing information of each activity becomes effectively random. The same algorithm trained and tested on randomized data has a performance that is not distinguishable from chance (Figure 4C), suggesting that all classification accuracy was derived from the timestamps of activities, and not differences in the abundance of certain activity types by chronotype.

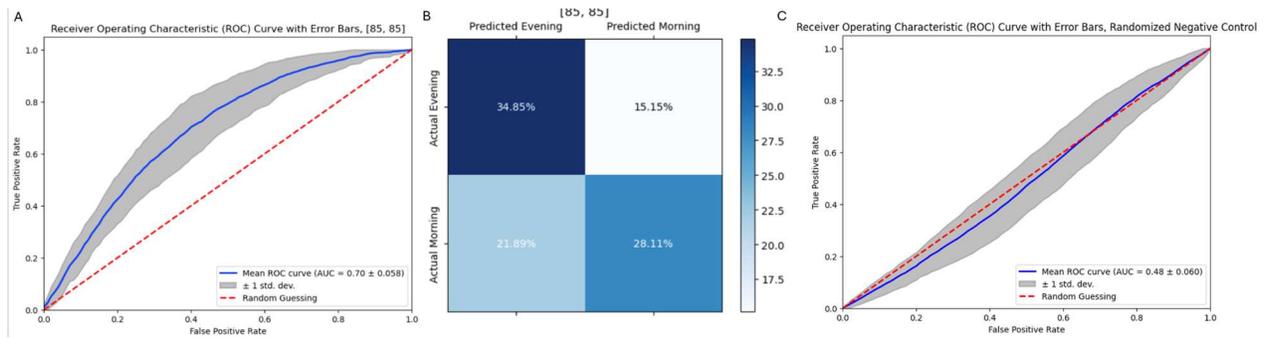

**Figure 4: Classifier performance and the importance of time of day**. **A)** The ROC curve evaluates the performance of the XGBoost binary classifier in distinguishing between morning and evening chronotypes based on users' (n=142) time series activity data. The mean ROC curve, shown in blue, had an Area Under the Curve (AUC) of 0.70 with a standard deviation of ±0.058. The shaded area along the blue curve represents the ±1 standard deviation range of the thirty runs conducted. **B)** The average confusion matrix depicts the classification performance of the XGBoost algorithm in predicting morningness and eveningness based on daily activity patterns. The matrix illustrates the distribution of true morning and evening chronotypes against their corresponding predicted classifications (n=142). Each cell represents the count of observations falling into the respective categories, providing insight into the algorithm's accuracy in distinguishing between morning and evening users. **C)** Negative control with values randomized with respect to time within each row independently. This ROC illustrates that without the implicit encoding of time within the data the model is unable to classify users better than random guessing (n=142).

# Discussion

Our analysis suggests digital real-world scheduling platforms, such as calendar or day planner apps, may provide adequate data to be useful in augmenting repeated and/or continuous chronotype assessment appropriate for certain interventions. Advantages of this method include reduced user burden to generate classifications, and the ability to monitor for longitudinal changes. Additionally, we found substantial variance in the timing across activity types as reported from real-world activities. This variability is consistent with our PCA findings of substantial overlap in activity timing, and our overall performance of 70% accuracy. Together, these findings support the hypothesis that there is not a persistent and clear separation of behaviors by chronotype in the real world. Instead, our findings suggest that in the real world there is substantial behavioral overlap between chronotypes, consistent with the presence of persistent social impositions on realized schedules. This is not shocking, as it might be expected given common class schedules and work schedules, for example. It highlights the need for multiple tools to assess chronotype and social jet lag across time, as this variability cannot be captured by one-time questionnaires or clinical measurements such as DLMO. It is interesting to note that our classifier performed at chance when timing of activity data were randomized. It is not a given that there would not be activity preference differences associated with chronotype.

For example, it could have hypothetically been the case that morning types got in exercise before work more easily than late types, especially if late types are also suffering exacerbations of sleep deprivation due to living with higher social jet lag burden [5,6]. If this were true, then the excess of exercise entries and of sleep entries in morning types would have provided some classification information even without regards to time. The lack of this evidence here suggests that either these differences were not present in this sample, or that they are small compared to the within-individual daily variability.

As the concept of chronotype evolves to include the phases and alignments of multiple oscillators per subject, and as methods evolve to track circadian disruptions (as in social jet lag) from different sources (e.g. meal timing may change independently from exercise timing or sleep timing), this variability may become an important source of information beyond what current chronotype questionnaires are designed to capture. Our findings therefore support the hypothesis that automated analysis of calendar app data provides an efficient means for extracting novel chronotypic information, but further research is needed to identify when this approach best serves a specific mental health intervention. Additionally, further research is needed to identify how chronotype information derived from novel, multimodal data sources corresponds with changes in internal physiological circadian mechanisms in different tissues.

## Limitations

Here we present only a study of the feasibility of automated chronotyping from real-world, multimodal calendar app data. Several limitations of the current approach are important to highlight. First, the data used are self-reported behavioral data, as compared to empirical biological measurements such as DLMO, cortisol, core body temperature, and genotype.[20–22] Whereas blood melatonin measurements are standardized across individuals,[23] definitions such as "exercise" and "relax" events may be understood and defined somewhat differently from user to user. The subjective limitation is not unique to calendar apps but applies to traditional survey instruments, such as MEQ and MCTQ as well.[24] We suggest that before such data or apps are used in any intervention, investigators should proactively educate users on intended definitions for each activity category, as part of their app onboarding experience, so the variation in subjective definitions for each activity type is minimized.

Another limitation is that we reduced the sample size to remove potentially artifactual data in the absence of certainty as to what was really artifactual. Furthermore, this model has been trained predominantly on users aged between 18 and 20, which may limit its generalizability. These data allowed our feasibility assessments, but any broader use of such classifiers requires the development of models appropriate across diverse populations. These limitations are in large part a limitation of the retrospective nature of the effort, and so future, prospective analyses or retrospective analyses over larger populations could further support the development of classification algorithms for which we confirm only feasibility here. Future efforts might also make sure of repeated sampling and classification to detect changes in classification trend over time. Given changes over time and classification accuracy of 70%, five independently classified weeks returning as "early type" in a row tells a different story than someone who is "early type" three times and "late type" two times over the same week. The latter might further be identified as at risk or unstable, and/or as appearing to respond to chronotype interventions. Appropriate statistical models should be developed in all cases, but the point not to lose is that these classifications can be used to generate signals over time, which can then be processed for additional information.

One challenge of building a classifier on such data is that our ground truth - gold standard MEQ chronotype survey scores, generate their classifications from questions about a free-running or idealized day.[25,26] Not only might such days be rare enough to limited relationship to everyday life, but they must be imagined in a context lacking sleep momentum or other sources of interference that likely impact any remembered holiday, for example. In our analysis, we analyzed weekends separately from weekdays to help account for professional and social obligations that may mask intrinsic preferences, or enable the classifier to learn despite the presence of social jet lag. However, more needs to be done to reconcile classifications from continuous, multimodal data showing real-world variability day to day to those classifications that are designed to ignore or aggregate such variability and multimodality. It remains to be seen when information from variable time series data adds more utility for a given intervention than retrospective surveys, and when the reverse is true. Care should be taken when planning interventions not to assume one form of assessment is universally more right than another, as both approaches depend on the trade-off of capturing or side-stepping real-world complexity and its impact on the manifestation of different daily and circadian rhythms. As we showed in our PCA example, people with different chronotypes as captured by the MEQ questionnaire show substantial overlap in some rhythmic outputs. We view this as demonstrating an opportunity to explore greater complexity in real-world daily rhythms than as an inability to make such data neatly fit into chronotype categories. Without a gold standard of direct measurement of circadian phase across tissues captured longitudinally (which is currently impossible), our findings additionally serve to highlight the uncertainty remaining in how chronotype can be definitively measured in the real world.

The current algorithm is limited in its successful assessment of morning and evening chronotypes. This study introduces the creation of a supervised XGBoost binary classifier to explore and validate the feasibility of this method for distinguishing between morning and evening chronotypes. Future iterations of the algorithm need to better accommodate intermediate chronotypes as well. Such efforts will entail developing methods capable of differentiating smaller variations in circadian rhythms and activity patterns, leading to a better understanding of individual chronobiological preferences. Additionally, leveraging more longitudinal data and considering additional factors, such as environmental influences and social interactions, could improve the classification process, providing more precise and personalized chronotype identification. Finally, comparison at a daily resolution between activity data like those used here, and biological assessments of different circadian phases across individuals' bodies would be ideal for mapping the relationships between these different kinds of rhythmic outputs. Such a study has not yet been attempted to our knowledge.

# Conclusions

Overall, the results indicate some potential in using multimodal app data to predict chronotypes, but they also highlight the significant challenges of applying binary labels to complex, multimodal behaviors. The overlap in activity patterns revealed both by the PCA and the complete loss of performance when randomizing activities by time suggests that real-world variability significantly influences daily schedules, complicating a discrete classification process. It is worth noting that activity tracking in aggregate as a statistical simplification resulted in classifications worse than chance. This highlights the need for multimodal data to improve the separation of populations, and suggests that high-dimensional and dynamical labeling schemes may more accurately reflect people's similarity to each other when such multimodal time series data are available. In the absence of such data, fixed labels of chronotype should be considered cautiously, as our findings align with other works suggesting a more complex and labile landscape of real-world timing.

The aim of this work was to test the feasibility of training an algorithm to classify MEQ score results based on calendar app data. We chose a supervised learning binary classifier , focusing initially on those with MEQ chronotypes either earlier or later than the "intermediate" category, and found substantial support for the feasibility of using calendar app data to gain information about chronotype. We also highlight the novel information that comes from calendar apps, as it provides a multimodal window into day-to-day variability of activity types and timing. While this complexity blurs the lines of clean and unambiguous classification, it also highlights the need, and provides a potential data source, for more research to understand how stability across different activity types interact with the underlying physiological timing systems to enable optimal well being on diverse individuals. Leveraging digital real-world calendaring data for automated and longitudinal chronotyping may present a useful new tool for scaling and broadening the implementation of chronotherapy. More research is needed to improve models, and also to expand the concept of chronotype and social jet lag to accommodate our multimodal lives.

# Author Contributions

PK, KA, SG, RM carried out data handling and analysis, and developed figures. KS, RK, and BS assisted with experimental design and analytic design. All authors contributed to the writing and editing of the manuscript.

## Conflict of Interest

BS and KS are advisors at Owaves, Inc., and RK is a founder; these three have a financial interest in Owaves Inc. They have no other financial interests to report. All other authors are contracted by Owaves for their contributions to this study and have no other conflicts of interest.

# References

1. Fischer D, Lombardi DA, Marucci-Wellman H, Roenneberg T. Chronotypes in the US – Influence of age and sex. *PLoS ONE*. 2017;12(6):e0178782. doi:10.1371/journal.pone.0178782
2. Druiven SJM, Riese H, Kamphuis J, et al. Chronotype changes with age; seven-year follow-up from the Netherlands study of depression and anxiety cohort. *J Affect Disord*. 2021;295:1118-1121. doi:10.1016/j.jad.2021.08.095
3. Shawa N, Rae DE, Roden LC. Impact of seasons on an individual's chronotype: current perspectives. *Nat Sci Sleep*. 2018;10:345-354. doi:10.2147/NSS.S158596
4. Roenneberg T, Kantermann T, Juda M, Vetter C, Allebrandt KV. Light and the human circadian clock. *Handb Exp Pharmacol*. 2013;(217):311-331. doi:10.1007/978-3-642-25950-0_13
5. Wittmann M, Dinich J, Merrow M, Roenneberg T. Social jetlag: misalignment of biological and social time. *Chronobiol Int*. 2006;23(1-2):497-509. doi:10.1080/07420520500545979
6. Smarr BL, Schirmer AE. 3.4 million real-world learning management system logins reveal the majority of students experience social jet lag correlated with decreased performance. *Sci Rep*. 2018;8(1):4793. doi:10.1038/s41598-018-23044-8
7. Zou H, Zhou H, Yan R, Yao Z, Lu Q. Chronotype, circadian rhythm, and psychiatric disorders: Recent evidence and potential mechanisms. *Front Neurosci*. 2022;16:811771. doi:10.3389/fnins.2022.811771
8. Fishbein AB, Knutson KL, Zee PC. Circadian disruption and human health. *J Clin Invest*. 131(19):e148286. doi:10.1172/JCI148286
9. Abbott SM, Malkani RG, Zee PC. Circadian disruption and human health: A bidirectional relationship. *Eur J Neurosci*. 2020;51(1):567-583. doi:10.1111/ejn.14298
10. Jonušaitė I, Sakalauskaitė-Juodeikienė E, Kizlaitienė R, et al. Chronotypes and their relationship with depression, anxiety, and fatigue among patients with multiple sclerosis in Vilnius, Lithuania. *Front Neurol*. 2023;14:1298258. doi:10.3389/fneur.2023.1298258
11. Walker WH, Walton JC, DeVries AC, Nelson RJ. Circadian rhythm disruption and mental health. *Transl Psychiatry*. 2020;10(1):1-13. doi:10.1038/s41398-020-0694-0
12. Codoñer-Franch P, Gombert M, Martínez-Raga J, Cenit MC. Circadian Disruption and Mental Health: The Chronotherapeutic Potential of Microbiome-Based and Dietary Strategies. *Int J Mol Sci*. 2023;24(8):7579. doi:10.3390/ijms24087579
13. Fishbein AB, Knutson KL, Zee PC. Circadian disruption and human health. *J Clin Invest*. 131(19):e148286. doi:10.1172/JCI148286
14. Reiter AM, Sargent C, Roach GD. Finding DLMO: estimating dim light melatonin onset from sleep markers derived from questionnaires, diaries and actigraphy. *Chronobiol Int*. 2020;37(9-10):1412-1424. doi:10.1080/07420528.2020.1809443
15. Klerman EB, Gershengorn HB, Duffy JF, Kronauer RE. Comparisons of the variability of three markers of the human circadian pacemaker. *J Biol Rhythms*. 2002;17(2):181-193. doi:10.1177/074873002129002474
16. Wan C, Mchill AW, Klerman EB, Sano A. Sensor-Based Estimation of Dim Light Melatonin Onset Using Features of Two Time Scales. *ACM Trans Comput Healthc*. 2021;2(3):1-15. doi:10.1145/3447516

17. *Read "Time-Use Measurement and Research: Report of a Workshop" at NAP.Edu*. doi:10.17226/9866
18. Roenneberg T. Twitter as a means to study temporal behaviour. *Curr Biol CB*. 2017;27(17):R830-R832. doi:10.1016/j.cub.2017.08.005
19. ten Thij M, Bathina K, Rutter LA, et al. Depression alters the circadian pattern of online activity. *Sci Rep*. 2020;10(1):17272. doi:10.1038/s41598-020-74314-3
20. Sládek M, Kudrnáčová Röschová M, Adámková V, Hamplová D, Sumová A. Chronotype assessment via a large scale socio-demographic survey favours yearlong Standard time over Daylight Saving Time in central Europe. *Sci Rep*. 2020;10(1):1419. doi:10.1038/s41598-020-58413-9
21. Allebrandt KV, Teder-Laving M, Kantermann T, et al. Chronotype and sleep duration: the influence of season of assessment. *Chronobiol Int*. 2014;31(5):731-740. doi:10.3109/07420528.2014.901347
22. Gershon A, Kaufmann CN, Depp CA, et al. Subjective Versus Objective Evening Chronotypes in Bipolar Disorder. *J Affect Disord*. 2018;225:342-349. doi:10.1016/j.jad.2017.08.055
23. de Almeida EA, Di Mascio P, Harumi T, et al. Measurement of melatonin in body fluids: Standards, protocols and procedures. *Childs Nerv Syst*. 2011;27(6):879-891. doi:10.1007/s00381-010-1278-8
24. Althubaiti A. Information bias in health research: definition, pitfalls, and adjustment methods. *J Multidiscip Healthc*. 2016;9:211-217. doi:10.2147/JMDH.S104807
25. Roenneberg T, Pilz LK, Zerbini G, Winnebeck EC. Chronotype and Social Jetlag: A (Self-) Critical Review. *Biology*. 2019;8(3):54. doi:10.3390/biology8030054
26. Reis C, Madeira SG, Lopes LV, Paiva T, Roenneberg T. Validation of the Portuguese Variant of the Munich Chronotype Questionnaire (MCTQPT). *Front Physiol*. 2020;11:795. doi:10.3389/fphys.2020.00795
27. Tosi, S. (2009). *Matplotlib for Python developers*. Packt Publishing Ltd.
28. J. D. Hunter, "Matplotlib: A 2D Graphics Environment", Computing in Science & Engineering, vol. 9, no. 3, pp. 90-95, 2007.
29. Salinas, D., Seeger, M., Klein, A., Perrone, V., Wistuba, M. & Archambeau, C.. (2022). Syne Tune: A Library for Large Scale Hyperparameter Tuning and Reproducible Research. *Proceedings of the First International Conference on Automated Machine Learning*, in *Proceedings of Machine Learning Research* 188:16/1-23 Available from https://proceedings.mlr.press/v188/salinas22a.html.
30. Pedregosa, F., Varoquaux, Ga"el, Gramfort, A., Michel, V., Thirion, B., Grisel, O., ... others. (2011). Scikit-learn: Machine learning in Python. *Journal of Machine Learning Research*, *12*(Oct), 2825–2830.
31. McKinney, W., & others. (2010). Data structures for statistical computing in python. In *Proceedings of the 9th Python in Science Conference* (Vol. 445, pp. 51–56).
32. Chen, T., & Guestrin, C. (2016). XGBoost: A Scalable Tree Boosting System. In *Proceedings of the 22nd ACM SIGKDD International Conference on Knowledge Discovery and Data Mining* (pp. 785–794). New York, NY, USA: ACM. https://doi.org/10.1145/2939672.2939785
33. Harris, C.R., Millman, K.J., van der Walt, S.J. et al. *Array programming with NumPy*. Nature 585, 357–362 (2020). DOI: 10.1038/s41586-020-2649-2.